\documentclass[10pt]{amsart}

\usepackage{amsfonts,amssymb,amsmath,bm}
\def\bbone{{\mathchoice {\rm 1\mskip-4mu l} {\rm 1\mskip-4mu l}
{\rm 1\mskip-4.5mu l} {\rm 1\mskip-5mu l}}}

\begin{document}

\ 

\smallskip
\begin{center}
\textbf{A massless quantum field theory over the p-adics}\\
\textsc{Abdelmalek Abdesselam}\\
(joint work with Ajay Chandra, Gianluca Guadagni)\\
\end{center}

\bigskip
\noindent
This talk is in three parts.
In Part 1, we briefly outine a general program for the rigorous study of scalar quantum field theory (QFT),
in the continuum. We use a probabilistic framework
in the spirit of Dobrushin~\cite{aa_Dobrushin}.
In Part 2, we explain that everything in Part 1 makes perfect sense when spacetime $\mathbb{R}^d$
is replaced by $\mathbb{Q}_p^d$.
Finally, in Part 3, we report on ongoing progress made, in collaboration with A.~Chandra and G.~Guadagni
(research funded by U.Va. and the NSF under grant DMS{\#}0907198),
on the $p$-adic analog of the massless model studied by Brydges, Mitter and Scoppola.

\section{Outline of a program for Euclidean QFT in the continuum}
The goal is to develop a mathematical theory which is a rigorous version of the methods
one finds in physics QFT textbooks (e.g., Ch.~8 and Ch.~10 of \cite{aa_ZinnJustin}).
We put the emphasis on Symanzik-Nelson positivity rather than reflection positivity.
The study of QFT thus becomes that of probability measures ${\rm d}\mu$ on the space of distributions
$S'(\mathbb{R}^d)$ with the cylinder $\sigma$-algebra.
We focus on measures which have finite moments and are invariant by translation (stationary processes) and
by the orthogonal group $O(d)$. One can also require self-similarity.
This program involves the following steps.

\noindent{\bf Step 0:}
It is to classify the self-similar Gaussian case. Let $\mathbf{S}_n(f_1,\ldots,f_n)=
\langle\phi(f_1)\cdots \phi(f_n)\rangle$
denote the moments of the measure ${\rm d}\mu$ under consideration. $\mathbf{S}_n$
can also be thought
of as an element of $S'(\mathbb{R}^{nd})$. In the (centered) Gaussian case, $\mathbf{S}_2$ contains all the information.
By translation invariance, $\mathbf{S}_2(x,y)=S_2(x-y)$ where $S_2\in S'(\mathbb{R}^d)$.
The classification reduces to that of $O(d)$-invariant distributions $S_2$ homogeneous of degree $-2[\phi]$
where $[\phi]$ is the scaling dimension of the field. For any $[\phi]\in\mathbb{R}$, there is a 1-dimensional
space of solutions.
Adding the positive-type condition entails $[\phi]\ge 0$. We will restrict the discussion to the range
$0<[\phi]<\frac{d}{2}$. One then has a simple expression in
both direct and Fourier space for the two-point function $\mathbf{S}_2(x,y)\sim |x-y|^{-2[\phi]}$,
$\widehat{S}_2(k)\sim |k|^{2[\phi]-d}$.
Note that this is only part of a bigger picture. One can handle zero-modes, e.g., by restricting $S(\mathbb{R}^d)$
using moment vanishing conditions~\cite{aa_Dobrushin}. See~\cite{aa_KangM} for $d=2$, $[\phi]=0$ which is pertinent
for conformal QFT. For work related to $d=1$, $[\phi]<0$, see the talk by J.~Unterberger.

\noindent{\bf Step 1:} Putting cut-offs. One replaces, e.g, the covariance $C=S_2$ by
$C_r(x)\sim\int_{L^r}^\infty \frac{{\rm d}\rho}{\rho} \rho^{-2[\phi]} u(\frac{x}{\rho})$
for some nice function $u$. Here $L$ is the renormalization group (RG) magnification ($L>1$ is an integer).
One also introduces a box $\Lambda_s$ of side length $L^s$.

\noindent{\bf Step 2:} Perturb the cut-off Gaussian ${\rm d}\mu_{C_r}(\phi)$ to get a new probability
measure ${\rm d}\nu_{r,s}(\phi)=\frac{1}{Z}\exp(-\widetilde{V}_{r,s}(\phi)){\rm d}\mu_{C_r}(\phi)$
where $\widetilde{V}_{r,s}(\phi)=\int_{\Lambda_s}{\rm d}^d x\ \{\ \widetilde{g}_r
:\phi^4:_{C_r}(x)+\ \widetilde{\mu}_r
\ :\phi^2:_{C_r}(x)+\cdots\}$.
Given a bare ansatz, i.e., the germ at $-\infty$ of a sequence $(\widetilde{g}_r,\widetilde{\mu}_r,\ldots)_{r\in
\mathbb{Z}}$, the key problem is to study the double limit ${\rm d}\nu_{r,s}\rightarrow{\rm d}\nu$
when $r\rightarrow -\infty$ and $s\rightarrow \infty$. One can either use the Bochner-Minlos Theorem
or a Hamburger moment reconstruction theorem with $n!$ growth for $\mathbf{S}_n$ in order to recover the wanted
QFT ${\rm d}\nu$.
To be interesting, ${\rm d}\nu$ should not be Gaussian. In fact, we would like a stronger notion of nontriviality
which begs the question: is there a good notion of Borchers class in this probabilistic setting?
Also of interest is the massless situation. Note that, for $\mu>0$, $(|k|^{d-2[\phi]}+\mu)^{-1}$
has large distance decay $|x|^{2[\phi]-2d}$ if $d-2[\phi]\notin 2\mathbb{N}$.
Thus, the appropriate definition of `massless' for general $[\phi]$ is the requirement of non $L^1$ rather
than power law decay for $S_2$. 

\noindent{\bf Step 3:} Composite fields and operator product expansion (OPE).
Borrowing our notation from the talk by S.~Hollands, we would like to define local field operators
$\mathcal{O}_A[\phi](x)$, e.g., renormalized versions of $\phi(x)^n$. After smearing by $f\in S(\mathbb{R}^d)$,
one would like $\phi\rightarrow \mathcal{O}_A[\phi](f)$
to be a function $S'(\mathbb{R}^d)\rightarrow\mathbb{C}$.
This typically fails if $[\phi]>0$. 
One should instead define $\phi\rightarrow \mathcal{O}_A[\phi](f)$ as a generalized function (or rather functional)
in the spirit of Hida's white noise calculus~\cite{aa_HidaKPS}. One needs a space $\mathcal{D}(S'(\mathbb{R}^d))$
of test functionals $F$ on
$S'(\mathbb{R}^d)$ which should at least contain monomials of the form $\phi(f_1)\cdots\phi(f_n)$.
Then $\mathcal{O}_A$ should be constructed as a linear map from $S(\mathbb{R}^d)$ to the dual space of generalized
functionals $\mathcal{D}'(S'(\mathbb{R}^d))$. The duality pairing is that given by the QFT/measure ${\rm d}\nu$.
Note that the correlations $\langle\mathcal{O}_A[\phi](f)\ F(\phi)\rangle=\langle\mathcal{O}_A[\phi](f)
\ \phi(f_1)\cdots\phi(f_n)\rangle$ make sense, in the free case, even at coinciding points.
Namely, this defines a distribution
on all of $\mathbb{R}^{(n+1)d}$. The functional $F$ corresponds to the spectator fields for the OPE.
In the case of a single operator insertion,
one can then follow the procedure explained in the talk by S.~Hollands, in order to study the singularities on
the diagonals and inductively define the operator products $\mathcal{O}_A$ from the corresponding short
distance asymptotics.
For the OPE with several operator insertions, one needs to define the mixed correlations at noncoinciding points, then
repeat the procedure.

\noindent{\bf Step 4:} Instead of perturbing, in Step 2, around a solution of Step 0,
one can also consider similar perturbations of nontrivial RG fixed points along relevant directions.

\section{The same over $\mathbb{Q}_p$}
The message here is that everything in Part 1 works perfectly if one considers
random fields $\phi:\mathbb{Q}_p^d\rightarrow\mathbb{R}$. Besides, the RG is much simpler and cleaner
than in the real case. Indeed, it reduces to the hierarchical RG.
For $p$ a prime number, the field $\mathbb{Q}_p$ is defined as the completion of the field $\mathbb{Q}$
with respect to the $p$-adic norm/absolute value
$|p^n\frac{a}{b}|_p=p^{-n}$, for $n,a,b\in\mathbb{Z}$ such that $b\neq 0$
and $p$ does not divide $ab$. A $p$-adic number $x\in \mathbb{Q}_p$
has a unique convergent representation $\sum_{j\in\mathbb{Z}} a_j p^j$, with only finitely many negative powers
of $p$, where the `digits' $a_j$ are in $\{0,1,\ldots,p-1\}$.
The polar part $\{x\}_p=\sum_{j<0} a_j p^j$ is a rational number. The valuation is given
by ${\rm val}_p(x)=\min\{j, a_j\neq 0\}$. The extention of the previous
norm is $|x|_p=p^{-{\rm val}_p(x)}$.
The unit ball $\mathbb{Z}_p=\{x\in \mathbb{Q}_p, |x|_p\le 1\}$ is a compact additive subgroup.
In dimension $d$, the norm of a point $x=(x_1,\ldots,x_d)$ in $\mathbb{Q}_p^d$ is defined by
$|x|=\max |x_i|_p$.
We take $L=p^l$ for the RG zooming ratio. The lattice of mesh $L^r$ is
given by $\mathbb{Q}_p^d/(L^{-r}\mathbb{Z}_p)^d$.
The big volume is $\Lambda_s=(L^{-s}\mathbb{Z}_p)^d=\{x, |x|\le L^s\}$.
For the space of test functions we take the Schwartz-Bruhat space $S(\mathbb{Q}_p^d)$ of locally constant
functions $f:\mathbb{Q}_p^d\rightarrow \mathbb{R}$ of compact support,
with the finest locally convex topology.
The Fourier transform is $\widehat{f}(k)=\int_{} f(x) {\rm e}^{-2i\pi\{x\cdot k\}_p} {\rm d}^d x$
where $x\cdot k=\sum x_i k_i$ and the additive Haar measure ${\rm d}^d x$ gives measure 1 to $\mathbb{Z}_p^d$.
The analog of $O(d)$ is the maximal compact subgroup $GL_d(\mathbb{Z}_p)$ of $GL_d(\mathbb{Q}_p)$, defined by fixing
the norm $|x|$. For Step 1, the cut-off covariance $C_r$ is obtained from
$C(x)\sim \sum_{j\in\mathbb{Z}} p^{-2j[\phi]}\bbone_{\mathbb{Z}_p^d}(p^j x)$
by imposing $j\ge rl$. The RG map corresponds to integrating over fluctuations
with covariance $C_0-C_1$.
With these modifications, Part 1 works in the $p$-adic setting too.
For other work on $p$-adic QFT see~\cite{aa_KochubeiS} and references therein.

\section{The $p$-adic BMS model}
The BMS model corresponds to $d=3$ and $[\phi]=\frac{3-\epsilon}{4}$ for some small positive
bifurcation parameter $\epsilon$. The bare ansatz only contains $\phi^4$ and $\phi^2$
couplings $\widetilde{g}_r$, $\widetilde{\mu}_r$.
We rescale to unit lattice $\widetilde{V}_{r,s}\rightarrow V_{r}^{(0)}$ and produce new bulk potentials
$V_{r}^{(0)}\rightarrow V_{r}^{(1)}\rightarrow\cdots$ by iterating the RG map (we suppressed $s$ in the notation).
Constructing a QFT morally amounts to establishing the transverse convergence criterion (TCC):
$\forall q\in\mathbb{Z}, \lim_{r\rightarrow -\infty} V_{r}^{(q-r)}$ exists (the effective theory
at log-scale $q$). This produces an ideal RG trajectory $(P_q)_{q\in \mathbb{Z}}$.
One conjectures that TCC $\Rightarrow \lim {\rm d}\nu_r={\rm d}\nu$ exists.
Together with A.~Chandra and G.~Guadagni  we adapted the proofs in~\cite{aa_BrydgesMS,aa_Abdesselam} to the
$p$-adic case and rigorously constructed
(in suitable Banach spaces) the nontrivial infrared (IR) fixed point, together with its stable and unstable manifolds.
We constructed ideal trajectories as well as established the TCC, starting from a bare ansatz,
for two massless theories: one which should be self-similar, at the IR fixed point, and another one which
joins the Gaussian and the IR fixed points.
Modulo the previous conjecture, we completed all previous steps except Step 3.
We are also making rapid progress towards proving this conjecture. This hinges on extending our
RG tools to nonuniform local perturbations of the massless Gaussian.
This should also help for Step 3.

\end{document}